\documentclass{ws-procs9x6}

\begin{document}

\title{Heavy quark studies with nuclear emulsions}

\author{G.~De Lellis, P.~Migliozzi, P.~Strolin}

\address{Universit\`a ``Federico II'' and INFN, Napoli, Italy, \\
Via Cintia Complesso Universitario \\
Monte Sant'Angelo, Edificio G-H, \\ 
80126 Napoli, Italy}


\maketitle

\section*{Foreword\footnote{by one of the authors (P.S.)} }

Although in 1964 CERN already had about 2000 staff members (though the
number of visitors, about 250, was much smaller than today), the human
environment maintained the spirit of friendship and cooperation
typical of smaller organisations. For instance, to come to CERN from
Geneva in the morning pedestrians simply had to stand at a given
corner in Rue de la Servette: the tacit understanding was that anybody
passing would give them a lift to CERN. I wonder when and why this
good habit was lost: too many people, too many cars?  

Indeed, it was also a very practical way to get to know each other. A
friend of mine freshly arrived at CERN was given a lift by "a" senior
physicist, whom he had not met before. Very frankly, he expressed his
worry because "they" were so late that morning and hoped that this
would not be noticed by the bosses. A few weeks later the CERN
Director General of that time, Viktor Weisskopf, started a new habit:
a regular meeting with newcomers, to get to know each other. My friend
was present at the first of these meetings. The Director General
started the meeting by explaining, through a story, how he had
understood the need for the initiative. The story of a newcomer to
whom he gave a lift one morning and, without recognising him as the
Director General, ...... That was CERN at that time and that was
Viktor Weisskopf: a great scientist who was very keen to communicate
with people at all levels and who was capable of doing it very
spontaneously and naturally.

So, I first met Roberto Salmeron when he gave me a lift to CERN. I was
immediately impressed by his culture and enthusiasm for physics and
society, as well as by his open, warm and human approach. My view of
Roberto has never changed.

It was only many years later, starting in the late seventies, that I
had the opportunity of closely collaborating with Roberto in the NA10
experiment at CERN. The experiment had been designed for the study of
muon pair production by quark-antiquark annihilation in the so-called
Drell-Yan process, using an intense pion beam to profit from the
valence antiquark content of the pions. Hence the experiment was aimed
at the study of hadron structure functions. When the beauty quark was
discovered at Fermilab, we realised that a special trigger for events
with three muons would have allowed us to carry out one of the first
measurements of the beauty production cross-section in hadronic
interactions. The motivation for writing the present paper on heavy
quark production stems from the memory of the splendid collaboration
with Roberto at that time.


\section{A historical introduction}
\label{introduction}
Photographic emulsions initiated accidentally their role in particle
physics more than 100 year ago, when H. Becquerel discovered natural
radioactivity by observing the blackening of photographic plates by
uranium salts~\cite{becque}. It was the beginning of a history which
led to the discovery of the ``new world'' of the elementary
particles.

Soon after the Second World War an intense R$\&$D carried out in
collaboration between the C. Powell group and Ilford led to the
development of the so called ``nuclear emulsions''. They are
characterised by very high sensitivity and grain uniformity, hence
they are capable of observing tracks of single particles with
submicrometric space resolution. This technique, specially suitable
for the observation of short lived particles, led to the discovery of
the pion~\cite{pion} in 1947 and to many important contributions to
the the development of particle physics. Many cosmic ray
experiments were carried out in the '50's and the '60's, contributing
among other observations to the discovery of the strange particles in
cosmic ray (see Ref.~\cite{Sacton:1998yf} and references therein).

In the meanwhile a new type of detector was proposed~\cite{ecc} and
then developed mainly in Japan, the so called Emulsion Cloud Chamber
(ECC). The ECC consists of a sandwich structure made of thick metal
plates (passive material) and thin emulsion layers (tracking
device). This detector was quite successful in cosmic ray experiments,
having the advantages of cost effectiveness and of particle
identification capabilities. Most of the detector mass consists of
metal plates, allowing for a substantial cost reduction ($\sim 1/100$)
compared with stacks of pure emulsions and allowing to reach higher
detector masses. In addition, the ECC allows the identification of
particles and the measurement of their kinematical parameter by
observing in detail specific ionisation, showering and multiple
Coulomb scattering.

\begin{figure}[ht]
\centerline{\epsfxsize=3.9in\epsfbox{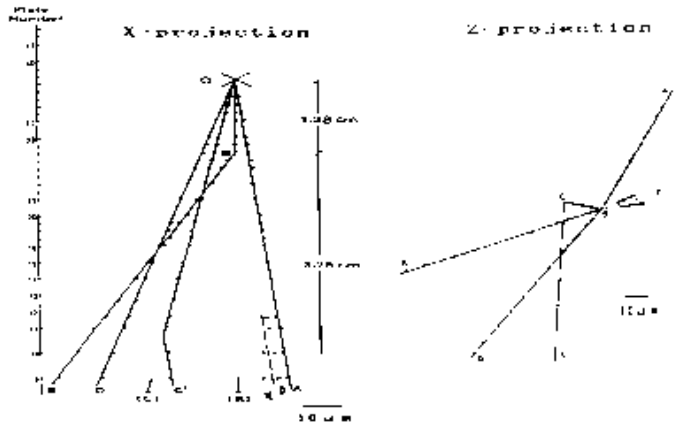}}   
\caption{Schematic drawing of the first evidence for the production
and decay of short lived ``$X$ particles'', now known as charmed
particles, in cosmic rays. \label{fig:firstcharm}}
\end{figure}

Among cosmic ray interactions detected in ECC, a very peculiar event
(see Fig.~\ref{fig:firstcharm}) was observed in
1971~\cite{Niu:1971xu}, three years before the discovery of of charm
with the observation of the
$J/\Psi$~\cite{Aubert:js}$^,$~\cite{Augustin:1974xw}$^,$~\cite{Bacci:1974za}.
The event is due to a neutral primary in the energy range of
10~TeV. The short track segment originating from the interaction
vertex was attributed to an ``$X$ particle'' of mass 2-3~GeV decaying
after $\sim10^{-13}$~s into a charged particle and a $\pi^0$
meson. The event was interpreted as the first example of the
associated production of a massive short-lived particle with a new
quantum number. After this observation, further examples of
$X$-particles (singly or pair produced) were then detected or dug out
from previous cosmic ray exposures, supporting the above
interpretation~\cite{Niu:ie}. By the time of the $J/\Psi$ discovery
and of the identification of the $X$-particle with a charmed meson,
about 20 $X$-particle decays had been observed by using the ECC
technique.

Thanks to the high space resolution of the ECC technique, both the
production and decay vertexes are reconstructed. Therefore, lifetime
studies started soon after the observation of the first $X$
particle. Just after the discovery of the $J/\Psi$, groups working
with the ECC could thus first report on the lifetime difference
between charged and neutral charmed particles. This early result
obtained by using cosmic rays was confirmed a decade later by the E531
experiment, on the Fermilab neutrino beam. In E531, about 120 charmed
particles were observed into an emulsion target and fully analysed by
using an electronic detector for a complete kinematical reconstruction
(see Ref.~\cite{Niu:ie} and references therein).

\begin{figure}[ht]
\centerline{\epsfxsize=3.9in\epsfbox{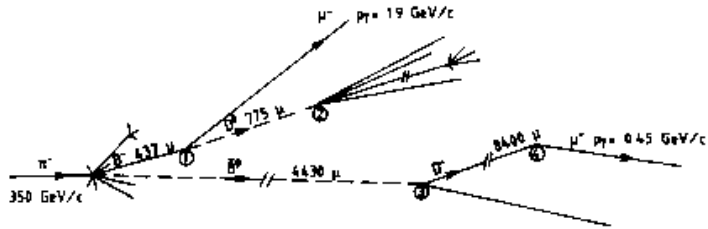}}   
\caption{Schematic drawing of the first hadro-produced $B\bar{B}$ pair
event observed in nuclear emulsions by WA75. \label{fig:beauty}}
\end{figure}

After the discovery in 1977~\cite{Herb:ek} of a new heavy quark,
i.e. of the beauty, experiments with nuclear emulsions aimed at the
direct observation of production and decay of beauty hadrons. A
successful search was first performed by the WA75 experiment at CERN,
using a $\pi^-$ beam of 350~GeV. In the WA75 event, shown in
Fig.~\ref{fig:beauty}, both beauty hadrons are observed to decay into
a charmed particle. Nine beauty events were later observed by the E653
experiment~\cite{Kodama:1992ff}.

\section{Heavy quarks, neutrinos and nuclear emulsions}
The high sensitivity and grain uniformity of nuclear emulsions, make
them capable of observing tracks of single particles with
submicrometric space resolution and therefore specially suitable for
the observation of short lived particles. One can say that nuclear
emulsions were the ancestor technique of heavy quark physics. The
present success of the emulsion technique is linked to the impressive
achievements in the development of the automatic scanning systems and
the conceptual design of the experiments. In the the latter the
adoption of the ``hybrid'' technique allow electronic detectors to
complement emulsions, to save time and to improve the event
reconstruction, as well as to complete the event reconstruction.

Neutrino and anti-neutrino induced charm-production is specially
interesting because it allow to isolate the strange-quark parton
distribution function and to study the transition to heavy quarks. In
particular, the understanding of the threshold behaviour associated
with the charm-production is critical for the extraction of
$\sin^2\theta_W$ from neutrino deep-inelastic data. Nowadays, this is
very important in relation with the recent measurement of
$\sin^2\theta_W$ performed by the neutrino experiment NuTeV, giving a
value of $\sin^2\theta_W$ which is more than $3\sigma$ away from the
average value measured at LEP~\cite{nutevsin}.

A better understanding of the inclusive charm-production cross-section
is also needed for the background determination of future experiments
aiming at the study of $\nu_\mu\leftrightarrow\nu_\tau$ oscillations.

Many electronic and bubble chamber experiments have studied indirectly
neutrino and anti-neutrino charm-production by looking at the presence
of two oppositely charged leptons in the final state. In the case of
neutrino scattering, the underlying process is a neutrino
charged-current (CC) interaction with an $s$ or $d$ quark, producing a
charm-quark that fragments into a charmed hadron. The charmed hadron
may decay semi-leptonically producing opposite sign di-leptons through
the process:

\begin{eqnarray}
\nu _\mu \;+\;{\rm N}\;\longrightarrow \;\mu ^{-}\!\! &+&\!c\;+\;{\rm X} \\
&&\!\!\hookrightarrow s\;+\;l ^{+}\;+\;\nu_l   \nonumber
\end{eqnarray}                                              

Analogously an anti-neutrino can interact with a $\bar{s}$ or
$\bar{d}$ anti-quark, producing a charm anti-quark that fragments into
a charmed hadron, again leading to a final state with two oppositely
charged leptons.

 This technique was first used in 1974~\cite{benvenuti} when neutrino
induced charm-production was discovered. Since then several
experiments~\cite{cdhs}$^,$~\cite{e616} have used this technique to
study charm-production.

Calorimetric experiments are characterised by a massive (iron) target
and a muon spectrometer to identify the muon and measure its
charge. Pion and kaon decays constitute the main background, in spite
of the high density and hence the short interaction length in the
target material. A further background reduction is obtained by
requiring a minimum momentum, typically $p_{\mu}>\mbox{5~GeV}$, for
each muon. The drawback of such a selection is that it prevents the
search for charm-production at low neutrino energies: for a typical
calorimetric experiment it is not possible to investigate energy
regions below 15~GeV, where the slow-rescaling threshold effect is
more important.

The main characteristic of a bubble chamber filled with a mixture of
heavy liquids (Ne-H$_2$, freon-propane) is its high efficiency in
identifying electrons. Therefore, they searched for charm-production
looking at $\mu^-e^+$ events. The low threshold,
$p_{e^+}>\mbox{0.3~GeV}$, combined with high statistics for
$E_\nu<\mbox{30~GeV}$, gives good sensitivity to the slow-rescaling
threshold behaviour. The main background sources are $\pi^0$ Dalitz
decays and $\bar{\nu}_e$CC interactions.

The main advantage of nuclear emulsion experiments studying charm
production is that, the charmed particle being identified through the
direct observation of its decay, the background is very low and very
loose kinematical cuts are applied. This gives a very good sensitivity
to the slow-rescaling threshold behaviour and consequently to the
charm-quark mass. Furthermore, inclusive and exclusive studies can be
performed and important results can thus be obtained in spite of the
statistics being much smaller than electronic experiments.

So far only two experiments, E531~\cite{Ushida:1988rt} and
CHORUS~\cite{Eskut:1997ar}, have searched for charm-production through
the direct identification of charm decays in emulsions, using neutrino
beams.

These experiments have a ``hybrid'' design, emulsions are used as
active targets. The main background for $D^0$ detection in emulsion
comes from $K^0$ and $\Lambda$ decays, and from neutron, $K^0$ and
$\Lambda$ interactions without any visible nuclear break-up at the
interaction point. For charged charmed-hadrons the main backgrounds
are $\pi$ and $K$ decays in flight, and the so called ``white kinks''
(hadron interaction without any visible track from nuclear break-up)
giving an apparent decay vertex. These backgrounds can be effectively
suppressed by applying a loose cut on the transverse momentum at the
decay vertex ($p_T>250~\mbox{MeV}$), so that they can be reduced to a
level of the order of $10^{-4}/CC$.

\section{An example of hybrid technique: the CHORUS detector}
\label{chorus}

The CHORUS detector~\cite{Eskut:1997ar} is a hybrid set-up that
combines a nuclear emulsion target with various electronic
detectors. The detector was designed to search for
$\nu_\mu\leftrightarrow\nu_\tau$ oscillations. Since charmed particles
and the $\tau$ lepton have similar lifetimes, the detector is equally
well suited for the observation of the production and decay of charmed
particles.

The nuclear emulsions act both as target for neutrino interactions and
as a high resolution detector, allowing three-dimensional
reconstruction of short-lived particles. The emulsion target has a
total mass of 770~kg and is segmented into four stacks, each
consisting of eight modules, each in turn composed of 36 plates with a
size of $36\times72~\mbox{cm}^2$. Each plate has a $90~\mu$m plastic
support coated on both sides with a $350~\mu$m emulsion
layer~\cite{Aoki:jh}. Each stack is followed by a set of scintillating
fibre tracker planes as electronic detector. Three interface emulsion
sheets act as interface with a $90~\mu$m emulsion layer on both sides
of an $800~\mu$m thick plastic base and by a set of scintillating
fibre tracker planes. The interface sheets and the fibre trackers
provide accurate predictions of particle trajectories into the
emulsion stack for the location of the vertex positions. The accuracy
of the fibre tracker prediction is about $150~\mu$m in position and
2~mrad in the track angle.

The additional electronic detectors downstream of the emulsion target
and associated trackers include a hadron spectrometer which measured
the bending of charged particles in an air-core magnet, a calorimeter
where the energy and direction of showers are measured and a muon
spectrometer which determines the charge and the momentum of muons.

\section{Nuclear emulsions analysis by automatic microscopes}
\label{scanning}



Emulsion analysis has been fully visual for several decades. Thus for
a long time emulsion were suitable only for low statistic experiments.
In the eighties, the development of semi-automatic systems opened a
new era characterised by the revival of nuclear emulsions. Nowadays
fully-automatic systems allow the reconstruction of several hundred
thousand interactions as in CHORUS. In the following we summarise the
original approach for automatic scanning developed in
Japan~\cite{aoki} with the so-called Track Selector (TS).

When observing an emulsion plate at the microscope, we observe a
``tomographic slice'' with a thickness of $\sim 5~\mu$m corresponding
to the of microscope focal depth. As emulsion plates are exposed
perpendicularly to the beam particle, in the tomographic image tracks
appear as grains. By raising or lowering the focal plane of the
microscope objective lens through the depth of the emulsion layer, the
three dimensional structure of tracks is reconstructed by connecting
aligned grains. The development of the automatic recognition system
for penetrating tracks has followed the model of human track
recognition. It is based on an integrated combination of mechanical
control and video image processing.

With an objective lens of magnification 50, the field of view is about
$120\times 150~\mu\mbox{m}^2$. Several tomographic (typically 16)
images are taken while changing the focal plane.  Each image is
suitably treated to reconstruct the grains and then tracks are formed
from consecutive grains.  This processing is carried out by hardware,
with a 3~Hz frequency.  A tracking efficiency of more than $98\%$ is
obtained for angles in the $[-0.4, 0.4~\mbox{rad}]$ range.

An upgrade of this system (UTS) with fast parallel processors working
at the same frequency has been performed in 1999~\cite{nakano}. A new
automatic scanning system (S-UTS) is currently under development. It
will increase the frequency up to 30~Hz. Key features of the S-UTS are
the high speed CCD camera with 3~kHz frame rate and a piezo-controlled
moving objective-lens, synchronised to the stage motion in order to
avoid stop-go of the microscope stage while taking images. Other
approaches for high speed emulsion analysis are being followed in
Europe. In recent years, the scanning speed has increased by one order
of magnitude every few years. The present aim is to reach a scanning
capability of 20~cm$^2$/s.



The impressive increase in scanning speed which has been achieved
allows scanning of large areas to locate events and to study their
topology. In addition, it makes it possible to perform in an automatic
way, measurements of kinematical parameters such as the momentum
measurement or electron identification through the measurement of the
multiple Coulomb scattering. In addition electron energy measurement
on the basis of the shower development can also be made.

\section{Charm-production in deep-inelastic neutrino interactions}
\label{charmtheo}

\subsection{Leading-order (LO) charm-production}
 
\begin{figure}[tbh]
\centerline{\epsfxsize=3.in\epsfbox{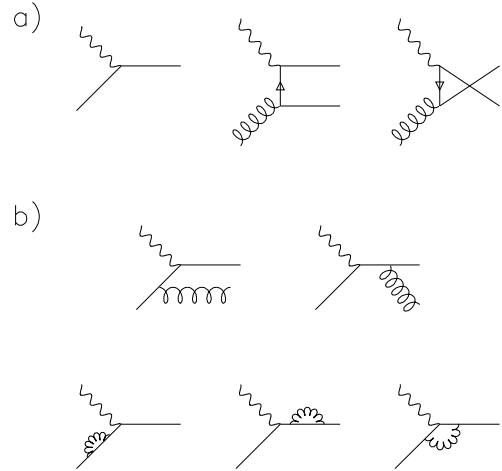}}   
\caption{Diagrams contributing to neutrino production, mediated by
charged gauge bosons, of charm-quark up to ${\mathcal
O}(\alpha_S)$. {\bf a)} The dominant diagrams: the leading order
quark-initiated diagram and the next-to-leading order t-channel and
u-channel gluon-initiated diagrams. {\bf b)} The radiative-gluon and
self-energy diagrams. \label{fig:cprod}}
\end{figure}
 
At leading order, only the Born diagram (the first diagram in
Fig.~\ref{fig:cprod}a)) is considered for charm-production and only
the $s$- and $d$-quarks of the nucleon are involved. In terms of the
quark distributions, for an isoscalar target the $c$ production
cross-section by neutrinos can be written as
 
\begin{eqnarray}                                                                \frac{d^2\sigma(\nu N\rightarrow c\mu^-)}{d\xi dy} & = & \frac{G_F^2ME}{\pi(1+Q^2/M^2_W)^2}\left\{ \left [ u(\xi,\mu^2)+d(\xi,\mu^2)\right ]\mid V_{cd}\mid^2 \right . \nonumber \\
& &\hspace{-2cm} \left . +2s(\xi,\mu^2)\mid
V_{cs}\mid^2\right\}\times\left [ \frac{1+R(\xi,\mu^2)}{1+\left
(\frac{2M\xi}{Q}\right )^2} \left ( 1-y-\frac{Mxy}{2E}+\frac{xy}{\xi}
\right) \right ]
\label{lo}
\end{eqnarray}
 
 
where $M$ is the nucleon mass, $E$ is the neutrino energy, $Q^2$ is
the negative square of the four-momentum transfer, $y$ is the
inelasticity, $R$ is a longitudinal structure function which accounts
for violation of the Callan-Gross relation, $\xi$ is the momentum
fraction of the struck quark and $\mu^2$ is the scale at which quark
momentum distributions are given. The dependence of the parton
distributions on the scale $\mu^2$ is specified by
QCD~\cite{Conrad:1997ne}. $\xi$ is related to the Bjorken scaling
variable $x$ through the relation
 
$$ \xi = x\left ( 1 +\frac{m_c^2}{Q^2}\right )\left ( 1 -
\frac{x^2M^2}{Q^2}\right)$$
 
For charm-production induced by $\bar{\nu}$, Eq.~\ref{lo} holds with
the substitution of quarks by anti-quarks.

The LO expression illustrates the sensitivity of the process to the
strange quark sea. Charm (anti-charm) production from scattering off
$d(\bar{d})$-quarks is Cabibbo suppressed. In the case of charm-quark
produced by neutrinos, approximately 50\% is due to scattering off
$s$-quarks, even though the $d$-quark content of the proton is
approximately ten times larger. In the case of anti-neutrino
scattering, where only sea $\bar{d}$-quarks contribute, roughly 90\%
is due to scattering off $\bar{s}$-quarks. Hence when neutrinos
interact the flavour of the struck quark is relatively well defined,
so that it IS possible to study specific flavour transitions.
 
\subsection{Next-to-leading order (NLO) corrections}

Because charm-production at LO is a process dominated by sea quark
contributions, the gluon-initiated contributions are expected to be
significant although they are nominally at
NLO~\cite{Conrad:1997ne}. The gluon distribution, which is an order of
magnitude larger than the sea quark distribution, compensates for the
extra power of $\alpha_s$ involved in the diagram (see
Fig.~\ref{fig:cprod}). Calculations including the NLO formalism have
recently become available~\cite{Kretzer:2001jn}.

\subsection{Hadronisation of charm-quarks to charmed hadrons}
In the previous Sections we discussed the charm-production at the
quark level, whereas the experiments only observe hadrons. The problem
of building hadrons (the so-called hadronisation) out of the quarks
emerging from a hard scattering process is characterised by a
relatively low scale, of the order of typical hadron masses, so that
perturbative QCD is not applicable.  

However, the factorisation theorem is not limited to the perturbative
region and is applicable to the hadronisation process in the same way
it is used when describing the initial state hadron. Just as the
parton distribution functions are universal, independent of the
scattering process through which they are probed, we may expect the
fragmentation description to be universal. In particular, one can
assume that quarks, emerging from $e^{+}e^{-} \rightarrow q\bar{q}$
scattering, form hadronic final states through processes which are
similar, if not identical, to the processes which build hadronic final
states in lepton-nucleon scattering.
 
The differential cross-section for charmed-hadron production by
neutrinos can be written as
 
\begin{eqnarray}
\frac{d\sigma(\nu N\rightarrow\mu^- C X)}{dxdydzdp_T^2} & =
&\frac{d\sigma(\nu N\rightarrow\mu^- c X)}{dxdy}\times\sum_h{f_h\times
D_c^h(z,p_T^2)}
\end{eqnarray}
 
Here, $D_c^h(z,p_T^2)$ is the probability distribution for the
charm-quark to fragment into a charmed hadron of type $h (= D^0, D^+,
D_s^+, \Lambda_c^+)$ carrying a fraction $z$ of the quark longitudinal
momentum and transverse momentum $p_T$ with respect to the quark
direction. The number $f_h$ is the mean multiplicity of the hadron $h$
in neutrino charm-production. A similar expression holds for anti-neutrinos.
 
Since only one $c$-quark is produced in a CC interaction, one can set
the normalisation conditions as
 
\begin{equation}
\int_0^1dz\int_0^\infty D_c^h(z,p_T^2)dp_T^2 = 1\,~~\mbox{and}~~~
\sum_h f_h = 1\,.
\end{equation}
 
Usually, $D_c^h(z,p_T^2)$ is written as
 
\begin{equation}
\label{bpar}
D_c^h(z,p_T^2) \propto D(z)\times e^{-bp_T^2} \end{equation}
 
where $D(z)$ is commonly parameterised either
as~\cite{Peterson:1982ak}
 
\begin{equation}
\label{pete}
D(z) \propto \frac{1}{z\big (
1-\frac{1}{z}-\frac{\varepsilon_P}{1-z}\big )}
\end{equation}
 
or as~\cite{Collins:1984ms}
 
\begin{equation}
\label{coll}
D(z) \propto \frac{\frac{1-z}{z}+\frac{\varepsilon_C(2-z)}{1-z}}{
1-\frac{1}{z}-\frac{\varepsilon_C}{1-z}}\times (1+z^2)
\end{equation}
 
Qualitatively, the shape of the $D(z)$ can be understood as follows.
The charm-quark emerges from the hard scattering process and is
relatively energetic. Attaching a light anti-quark $\bar{q}$ to it to
form a charmed meson decelerates the heavy quark only slightly, so
that we expect the variable $z$ to be peaked toward 1.  The
$\varepsilon$ parameter is different for different mesons, but it
can be expected to scale as $1/m_{Q}^{2}$, with $m_{Q}$ the mass of
the heavy quark.
 
Finally, we stress that all the parameters discussed above ($f_h$,
$b$, $\varepsilon_P$, $\varepsilon_C$) cannot be predicted
theoretically and must be determined experimentally.

\section{Physics results and perspectives}
\label{physres}
\subsection{Cross-section measurements}
\label{cross}

Inclusive charm-production cross-sections can be only measured in
emulsion experiments, while exclusive measurements can also be
performed by electronic experiments like
NOMAD~\cite{Astier:2001ri}. As already mentioned, another important
advantage of the emulsion technique is the possibility to observe
charm-production down to very low energies with a very low background,
allowing the investigation of the threshold behaviour and the
extraction of the strange-quark parton distribution function at low
$Q^2$.

The total charmed-particle production inclusive cross-section relative
to CC interactions measured by the E531 experiment is: $\sigma(\nu_\mu
N\rightarrow\mu^- cX)/\sigma(\nu_\mu N\rightarrow
\mu^-X)=4.9^{+0.7}_{-0.6}\%$.

The inclusive $D^0$ production cross-section times its branching ratio
into charged particles has been measured by the CHORUS
experiment~\cite{Kayis-Topaksu:2002mb} and it amounts to
$\sigma(\nu_\mu N\rightarrow D^0\mu^-X)/\sigma(\nu_\mu N\rightarrow
\mu^-X)=(1.99\pm0.13\pm0.17)\%$.

The total charmed-particle production inclusive cross-section in
anti-neutrino induced events has never been measured. Its measurement
is currently in progress in the CHORUS experiment.
 
\subsection{Charm-hadronisation studies}
\label{hadro}

\subsubsection{Charmed fractions and semi-muonic branching ratio}
\label{bmu}

So far, the charmed fractions $f_h$ have been directly measured only
by the E531 experiment~\cite{Ushida:1988ru}. A bias was later detected
in the extraction of the charmed fractions by E531 and the data was
refit with the bias removed~\cite{Bolton:1997pq}.

When comparing results from neutrino measurements with $e^+e^-$
experiments with similar kinematics, one should account for the fact
that neutrinos have the peculiarity to undergo not only deep-inelastic
interactions, but also diffractive and quasi-elastic charm-production
(see Section~\ref{lowmult}). The results of the refit to E531 data
together with a prediction based on $e^+e^-$ experiments plus the
corrections for neutrino ``peculiarities''~\cite{charfrac} are
consistent with the assumption that the charm quark hadronises
independently of the process through which it has been produced.
 

Recently, a direct measurement of the semileptonic branching ratio
${B}_\mu$ has been performed by CHORUS using about 1000 charm events
reconstructed in nuclear emulsions~\cite{VandeVyver:2002uu}. Out of
these, ($88\pm10\pm8$) dimuon events have been reconstructed, which
correspond to

$${B}_\mu = (9.3\pm0.9\pm0.9)\%$$

in agreement with previous indirect measurements by electronic
experiments. A precise determination of $B_\mu$ is particularly
important for a high precision extraction of the CKM element matrix
$V_{cd}$ from dimuon experiments~\cite{Conrad:1997ne}.

The CHORUS experiment is now analysing a charm sample ten times larger
than the E531 one, with the aim to both measure accurately the charmed
fractions and reduce the error on $B_\mu$.

\subsubsection{Transverse momentum distribution}



The transverse momentum squared ($p_T^2$) distribution of charmed
hadrons with respect to the direction of the hadronic system has been
measured by E531~\cite{Ushida:1988ru} and, using electronic detectors,
by NOMAD~\cite{Astier:2001ri}. The $p_T^2$ distribution from E531 and
NOMAD are consistent. The averaged value of the exponential slope
parameter $b$ (see Eq.~\ref{bpar}) is $3.31\pm0.27~\mbox{GeV}^{-2}$.

 CHORUS is currently analysing about 1000 events with a charmed hadron
in the final state and is expected to provide a new insight in the
understanding of the $p_T^2$ distribution of charmed hadrons.

\subsubsection{Fragmentation studies}

The parameter $\varepsilon$ which characterises the fragmentation
functions (see Eqs.~\ref{pete} and~\ref{coll}) of heavy quarks can be
determined by using two different approaches:

\begin{itemize}

\item Direct measurements: the $z$ distribution is reconstructed and
fitted in order to extract the parameter $\varepsilon$. Such analysis
has been performed by E531 and NOMAD. It is currently in progress in
CHORUS;

\item Indirect measurements: $\varepsilon$ is left as one of a free
parameters of the fit to dimuon data.  For details of dimuon
analyses performed by the electronic experiments CDHS, CCFR, CHARM II
and NuTeV we refer to~\cite{dimuon}.
\end{itemize}

The available results from both approaches are summarised in
Table~\ref{tab:eps}, together with the results from $e^+e^-$
experiments (CLEO and ARGUS) at $\sqrt{s}=10.55$~GeV.

\begin{table}[htb]
\tbl{Summary of all available determinations of $\varepsilon$ from
neutrino experiments. For comparison results from $e^+e^-$
(CLEO+ARGUS) experiments are also given. \vspace*{1pt}} {\footnotesize
\begin{tabular}{|c|r|r|r|}
\hline
{} &{} &{} &{}\\[-1.5ex]
Collaboration & $\varepsilon_P$ & $\varepsilon_C$ & Comments \\[1ex]
\hline
{} &{} &{} &{}\\[-1.5ex]
E531~\cite{Ushida:1988ru} & $0.076\pm0.014$ & --- & All charmed hadrons are used\\[1ex]
NOMAD~\cite{Astier:2001ri} & $0.075\pm0.046$ &  $0.13\pm0.14$ & Only $D^{*+}$ are used \\[1ex]
CDHS~\cite{Abramowicz:1982zr} & $[0.02\div0.14]$ & --- &  \\[1ex]
CCFR (LO)~\cite{Rabinowitz:1993xx} & $0.22\pm0.05$ &  $0.88\pm0.12$ & All charmed hadrons are used \\[1ex]
CHARM II~\cite{Vilain:1998uw} & $0.072\pm0.017$ & --- & All charmed hadrons are used\\[1ex]
CLEO+ARGUS~\cite{pdg} & $0.14\pm0.01$ & --- &Only $D^0$ are used\\[1ex]
CLEO+ARGUS~\cite{biebel} & $0.156\pm0.022$ & --- &Only $D^+$ are used\\[1ex]
CLEO+ARGUS~\cite{biebel} & $0.10\pm0.02$ & --- &Only $D_s$ are used\\[1ex]
CLEO+ARGUS~\cite{pdg} & $0.25\pm0.03$ & --- &Only $\Lambda_c$ are used\\[1ex]
CLEO+ARGUS~\cite{pdg} & $0.078\pm0.008$ & --- &Only $D^{*+}$ are used\\[1ex]
\hline
\end{tabular}\label{tab:eps} }
\vspace*{-13pt}
\end{table}
 
Given the large charm statistics of the CHORUS experiment, it will be
possible in the near future to study with high accuracy fragmentation
parameters as a function of the charmed hadron type and compare them
with the results obtained by $e^+e^-$ experiments.

\section{Low multiplicity charm-production}
\label{lowmult}
We classify as ``low multiplicity'' those processes (diffractive
$D_s^{(*)}$ and quasi-elastic (QE) charm-production) with at most two
particles produced at the primary vertex, besides the charmed hadron.

Quasi-elastic charm-production, is particularly interesting for the
measurement of the absolute branching ratio (BR) of the
$\Lambda_c$~\cite{Migliozzi:1999ca}. In fact, events quasi-elastically
produced have a peculiar topology which allows to define an almost
pure sample of $\Lambda_c$ and consequently to extract a model
independent measurement of the absolute BR of the
$\Lambda_c$~\cite{Migliozzi:1999ca}. In the past this process has been
studied with a small statistics (less than 10~events total world
statistics)~\cite{Mangano:2001mj}. Presently, the CHORUS experiment is
performing a dedicated search and several events consistent with a QE
topology have been already measured. With the final statistics of a
few hundred events, CHORUS will measure for the first time both the
differential cross-section of the process and the absolute BR of the
$\Lambda_c$.

\section{Associated charm-production in neutral-current and charged-current interactions}
\label{asso}
Associated charm-production in NC interactions proceeds through a
gluon-boson fusion process. Only one event has been ``directly''
observed by the E531 experiment~\cite{Ushida:1988rt}, while an
indirect observation has been recently published by the NuTeV
experiment~\cite{Alton:2000ze}. Both experiments measured a rate of
the order of $10^{-3}$, normalised to CC neutrino interactions.

In CC interactions, charm-anti charm pairs originate from a gluon
emitted through the bremsstrahlung off a light quark. In the past,
indirect evidence for this process was obtained by studying trimuon
and same-sign dimuon events~\cite{Sandler:wj}. The puzzle was that the
observed rate of trimuons and same-sign dimuons was larger than
theoretical predictions~\cite{Hagiwara:1980nu} by more than one order
of magnitude. Although, given the small statistics and the
uncertainties related to the background subtraction, the result was
not conclusive, the observed large discrepancy motivates further
theoretical and experimental studies. Recently a direct search for
this process has started in the emulsions of the CHORUS experiment. So
far the observation of one event~\cite{ccbar} has been reported.

%
%

\section{Conclusions}

Emulsions have started particle physics with the discovery of natural
radioactivity by Becquerel in 1896. The development of the ``nuclear
emulsions'' made it possible to detect tracks of single particle and
to perform detailed measurements of their interactions. The discovery
of the pion in 1947 was the first, spectacular demonstration of their
unique features for the direct observation of the production and decay
of short-lived particles, with negligible or very low background. In
particular, these features are now exploited for studies of heavy
quark physics in experiments where nuclear emulsions are combined with
electronic detectors and profit is taken of the remarkable
technological progress in automated analysis. In these experiments,
neutrinos provide a selective probe for specific quark
flavors. Interesting results on charm production and decay are
expected in the very near future, well in time to celebrate with
physics results the 90th anniversary of Roberto Salmeron.

\end{document}